\documentclass[12pt,a4paper]{article}
\usepackage{jheppub}
\usepackage{graphicx}
\usepackage{subcaption}
\usepackage[titletoc]{appendix}
\usepackage{xspace}
\usepackage{hyperref}
\usepackage[utf8]{inputenc}
\usepackage[backend=bibtex,style=numeric-comp,sorting=none]{biblatex}
\begin{document}

\title{HL-LHC Computing Review Stage 2, \\ Common Software Projects:\\
Data Science Tools for Analysis}
\author[1]{Jim Pivarski~(editor),}
\author[2]{Eduardo Rodrigues~(editor),}
\author[3]{Kevin Pedro~(editor),}
\author[4]{Oksana Shadura,}
\author[5]{Ben Krikler,}
\author[6]{Graeme A. Stewart}
\author{\\for the HSF PyHEP Working Group}

\affiliation[1]{Princeton University, Princeton NJ, USA}
\affiliation[2]{Oliver Lodge Laboratory, University of Liverpool, Liverpool, UK}
\affiliation[3]{Fermilab, Batavia IL, USA}
\affiliation[4]{University of Nebraska-Lincoln, Lincoln NE, USA}
\affiliation[5]{University of Bristol, Bristol, UK}
\affiliation[6]{CERN, Geneva, Switzerland}

\maketitle

\section{Description and relevance for the HL-LHC}

Unlike other documents prepared for the \textit{HL-LHC Computing Review Stage 2}, this ``Data Science Tools for Analysis'' document is not an experiment or software product with centralized management. The title describes a collection of interrelated small projects, developed by a network of teams or individuals in different organizations and communities. It is relevant to HL-LHC planning because many of the developers and users of this software are LHC physicists, particularly the Ph.D.\ students and postdocs involved in the actual delivery of LHC analyses.

We use the word ``data science'' for two reasons. The first reason is that most of the foundational software underlying the ecosystem and work discussed in this document were developed by cross-disciplinary scientists (i.e.\ not just HEP) and by statistician-programmers in industry: business intelligence, finance, web analytics, and other fields that require timely calculations on large datasets---collectively known as ``data science.'' The second reason is that the culture of distributed software development that is taking hold in HEP derives from exposure to a similar culture in the data science industry. The norms are changing: it has become more common for developers to contribute to a variety of related projects, not just their own, and to value fine-grained packaging in an interoperable ecosystem.

Although many of the data-science-based projects in HEP are self-motivated---physicists spontaneously writing software to fulfill the needs of their own analyses---we can't simply ignore such developments and assume that it will all work out in the end. Since the rise of open-source software in the late 1990's/early 2000's, the decentralized nature of its development has been in tension with business needs for stability, reliable planning, and influence over the direction of development---just as we need to prepare for the HL-LHC over a decade in advance. In fact, it's somewhat surprising that this ``chaotic'' model of software development works at all; but, crucially, 20~years of experience has shown that it does. A ``tragedy of the commons'' does not apply because software is not a scarce resource like grazing pasture---it can't be used up by a corporation taking it without recompense. Moreover, communal software development provides its own incentive for support. Software products are moving targets: if stakeholders do not contribute their own modifications to the ever-changing community version, they risk having to maintain an unmergeable fork. All of this has led to the remarkable situation in which billion dollar companies now contribute to data analysis software that is available for free.

The relevance of this to the HL-LHC is that our scientific goals are also a specialized interest, and we can either influence these shared software products to meet our needs or we will have to maintain our own software stacks. To an extent, young physicists (i.e., graduate students and postdocs) are making this decision for us by adopting data science software in their analyses and contributing back to it, as well as developing their own packages to satisfy HEP-specific needs. This has been a clear trend in the past 3--4~years, with community support and the emergence of community projects. We can support this work by connecting it to (not absorbing it into) established HEP products like ROOT~\cite{citeulike:363715} and the LHC experiment frameworks, and by mediating communication so that these small-granularity software products can each find their niche, and only one per niche. We can also steer the grassroots efforts in a forward-looking direction: physicists may be developing tools that suit their own analyses now, but will they scale to HL-LHC luminosities?

Data science software in HEP is a side job for many of its developers, for whom a physics analysis or a thesis is the top priority. That makes it particularly vulnerable---the immediate need to get a physics project done quickly very often outweighs the future need to get projects like it done in a better way. Git history is littered with abandoned projects such as these. On the other hand, the will to develop such software is already, and strongly, present in the community. We do not have to convince anybody to do it; we just need to ensure that it grows in the right direction.

\section{The HEP analysis software landscape is changing}

Although we can find evidence of physicists attempting to use tools such as NumPy, SciPy, Matplotlib, Pandas, Hadoop, and Spark over the past 15--20 years, these data science tools have only become a major part of the HEP analysis ecosystem in the past 4--5 years. Causes for this change are hard to identify conclusively because it is a bottom-up movement, but the following are likely influences:

\begin{itemize}
\item Externally, data analysis tools have consolidated on Python and array-oriented programming. The landscape had previously been more fractured: in the early 2000's, novel statistical techniques were implemented in R and big datasets were managed by Java, but in the past 5~years, Python has become the common language for both. Because of these changes in the industry, young physicists are much more likely to see Python in university courses and code examples on the web, and a career path that involves Python expertise is far more attractive than one without. Also, ``the industry'' is orders of magnitude larger than HEP, and many physics students are at least considering careers outside of HEP. A significant fraction of the newcomers to our field have prior experience with Python and have good reasons to leverage it and to want more.

\item Internally, organizations within HEP have mediated communications about data science software, helping physicist-developers find each other, reduce duplication, and build on each others' tools. These organizations have legitimized the idea of using data science software in HEP, and they are actively supporting its development, especially for the foundational components that enable specialized tools.
\end{itemize}

The first of these organizations, the \href{https://hepsoftwarefoundation.org/}{HEP Software Foundation (HSF)}~\cite{HSF}, was conceived in a community event at SLAC~\cite{hsf-workshop-slac2015} in 2015 and cohered in the milestone \href{https://hepsoftwarefoundation.org/organization/cwp.html}{Community Whitepaper endeavor}~\cite{Albrecht2019} of 2017. The HSF's primary mission is communication: helping project developers find each other to work toward a common vision, as well as providing fora in which new projects are conceived, particularly across experiment boundaries. The HSF does not fund projects or direct them. The \href{https://hepsoftwarefoundation.org/workinggroups/dataanalysis.html}{Data Analysis Working Group (DAWG)} hosts monthly meetings on data analysis software, giving physicist-developers a place to showcase their work and find users and collaborators.

The \href{https://hepsoftwarefoundation.org/workinggroups/pyhep.html}{``Python in HEP'' Working Group (PyHEP)} brings together a community of developers and users of Python in Particle Physics, with the aim of improving the sharing of knowledge and expertise. It embraces the broad community, from HEP to the Astroparticle and Intensity Frontier communities. Activities-wise it hosts large, annual workshops on both domain-specific and infrastructure libraries, as well as Python Module of the Month meetings to focus on data science and HEP domain-specific packages of interest.

\href{http://diana-hep.org/}{DIANA/HEP}~\cite{DIANA/HEP} and its much larger successor, \href{https://iris-hep.org/}{IRIS-HEP}~\cite{IRIS-HEP}, were funded by the NSF in 2015 and 2018, respectively, to support the development of HEP software (a)~financially, by funding software developers, (b)~as an intellectual hub for sharing knowledge and fostering connections, and (c)~through training and education. The directly funded software products include Awkward Array, Hist, pyhf, recast, ServiceX, SkyhookDM, Uproot, and Vector, all of which have substantially influenced the Python/data science ecosystem in HEP. Weekly DIANA/HEP and IRIS-HEP topical meetings are held by physicists and occasionally draw data scientists from industry to share tools and techniques, and IRIS-HEP hosts cross-experiment software tutorials, as well as the annual \href{http://codas-hep.org/}{CoDaS-HEP school}.

Other national projects, such as \href{https://gtr.ukri.org/projects?ref=ST\%2FV002562\%2F1}{SWIFT-HEP} in the U.K., are fulfilling a similar role to IRIS-HEP in the U.S.;
SWIFT-HEP started in Spring 2021.

Less formally, the \href{https://scikit-hep.org/}{Scikit-HEP} community project~\cite{Rodrigues:2020syo}, initiated  by several HEP physicists from several experiments in Autumn 2016, has focused efforts to build a Pythonic toolset ecosystem for analysis in HEP---the first such community project---by providing a common goal, brand, high standards of quality packaging, and a web presence. It grew dramatically, providing many of the packages listed above, with strong community adoption. The developers of software projects like \href{https://coffeateam.github.io/coffea/}{Coffea} (U.S. project), \href{https://fast-hep.web.cern.ch/fast-hep/}{FAST-HEP} (U.K. project) and \href{https://zfit.readthedocs.io/en/latest/}{zfit} (Swiss project), also have a strong influence on the ecosystem, since each widely-used software package accretes a planetary system of related tools. Coffea~\cite{Smith:2020pxs}, a Fermilab project that glues together foundational components and adds whatever is missing for analysis, has been particularly influential,
albeit mostly within CMS. It can be credited with the initial promotion of Awkward Array~\cite{Pivarski:2020txo} and it acts as a rapid testing ground for components that eventually spin off into their own infrastructure components, such as Hist (for histograms) and Vector (for Lorentz vectors).

\begin{figure}
\centering
\includegraphics[width=0.95\linewidth]{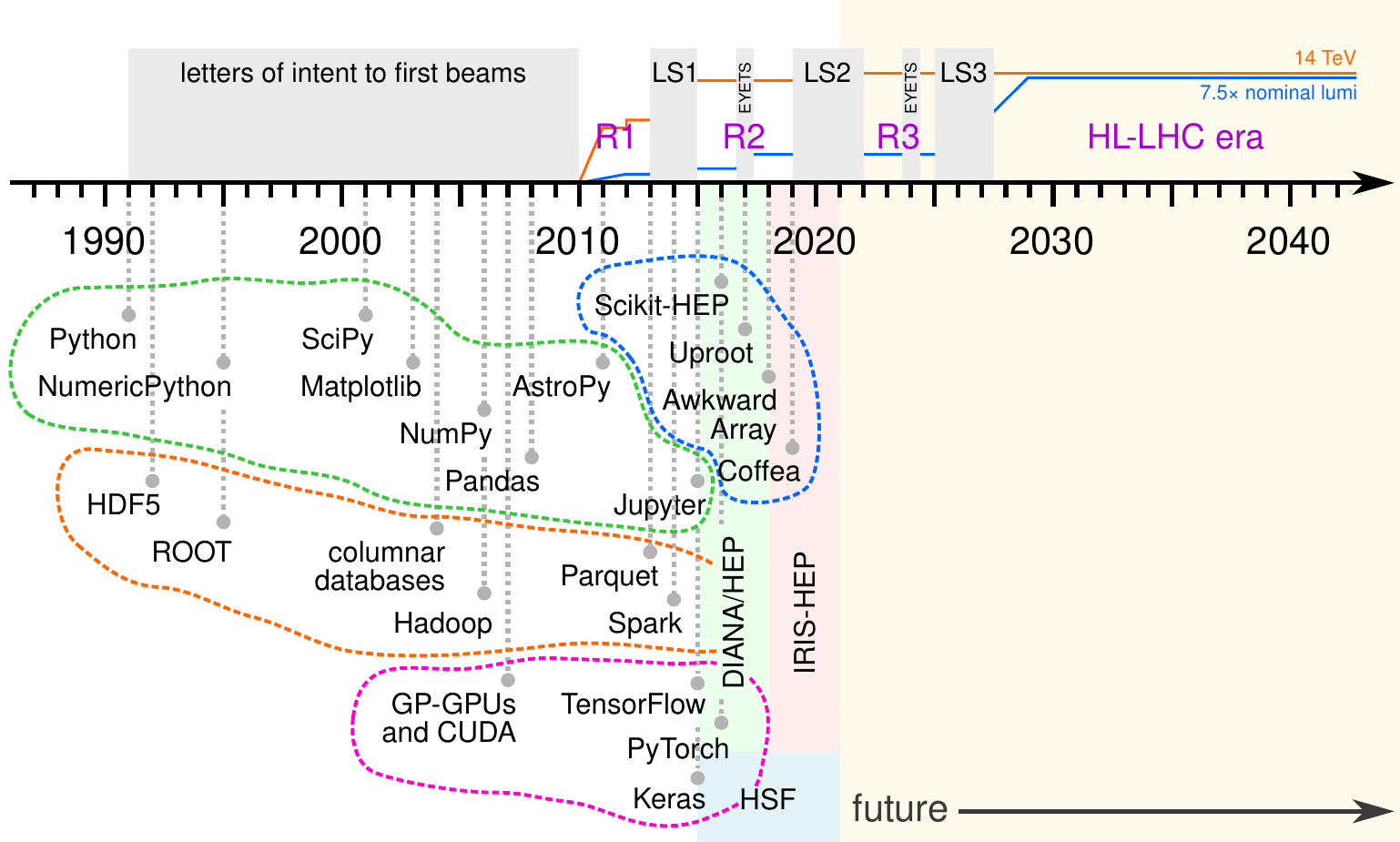}

\caption{Timeline of developments in scientific Python (dashed green outline), big data (orange), GPUs and machine learning (magenta), and data science in HEP (blue), overlaid on the LHC, HL-LHC, HSF, DIANA/HEP, and IRIS-HEP timelines. \label{fig:hllhc-python-timeline-paper}}
\end{figure}

The big picture of developments in data science software, HEP software, and the LHC/HL-LHC timeline are depicted together in Figure~\ref{fig:hllhc-python-timeline-paper}. Major developments in data science---scientific Python, big data, GPUs and machine learning---roughly coincide with the development and first run of the LHC, from the late 1990's to 2015. The HSF, DIANA/HEP, and IRIS-HEP were active after 2015, and these organizations fostered the development of data science-oriented software in HEP.

\begin{figure}
\centering
\includegraphics[width=0.8\linewidth]{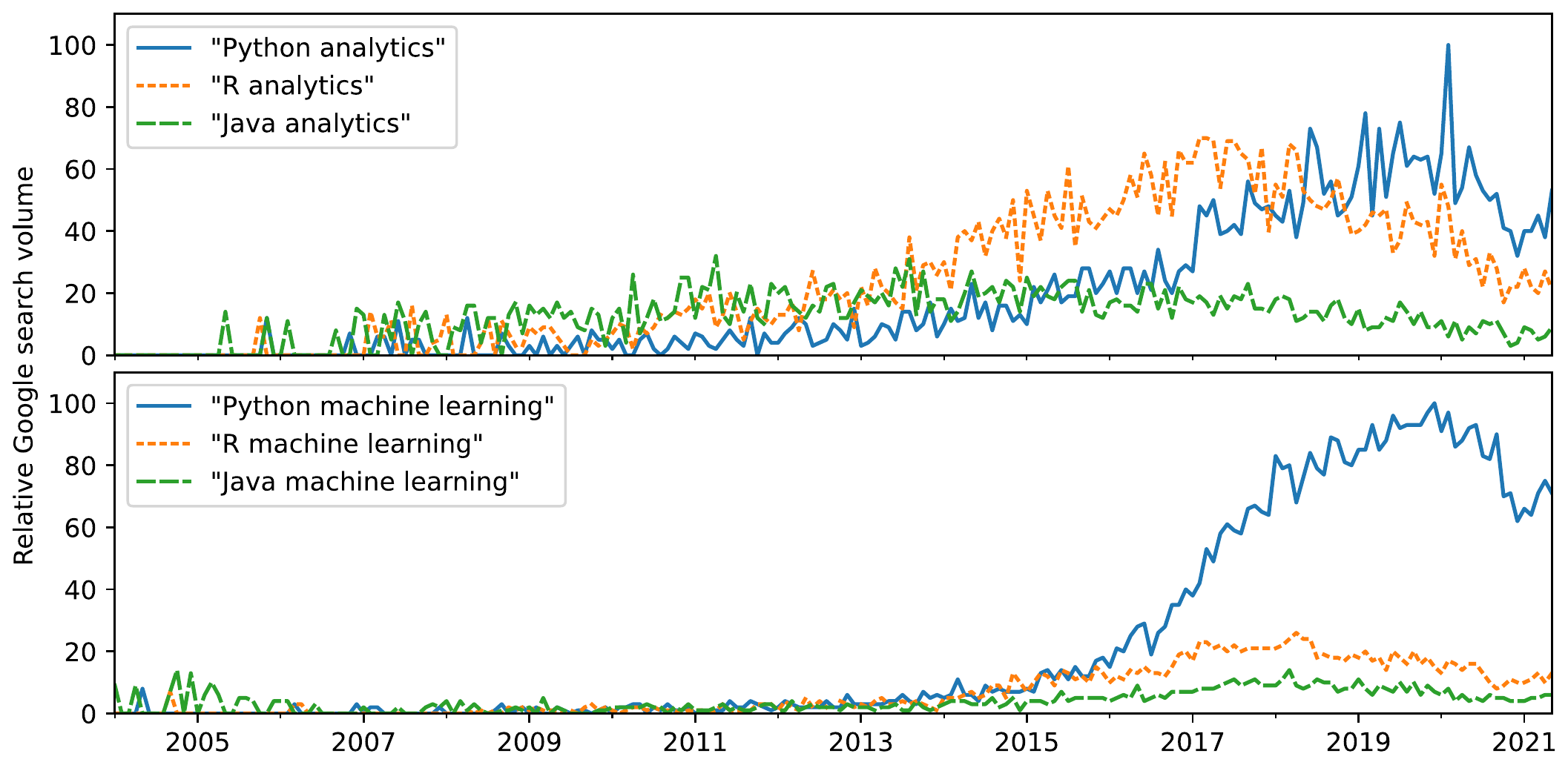}

\caption{Relative Google search volume (from \href{https://trends.google.com/}{trends.google.com}) for ``Python,'' ``R,'' and ``Java'' in the same search string with ``analytics'' (top) and ``machine learning'' (bottom). \label{fig:analytics-by-language}}
\end{figure}

Originally, data science in industry was split between R for statistical techniques, Java for scale-out (particularly Hadoop and Spark), and Python. Figure~\ref{fig:analytics-by-language} shows the ascendancy of Python as the language of data analytics, as measured by Google search terms. By 2018, ``Python'' was more frequently searched with ``analytics'' than ``R'' or ``Java,'' and machine learning, which entered a renaissance in 2015, was Python-focused from the start.

\begin{figure}
\centering
\includegraphics[width=0.8\linewidth]{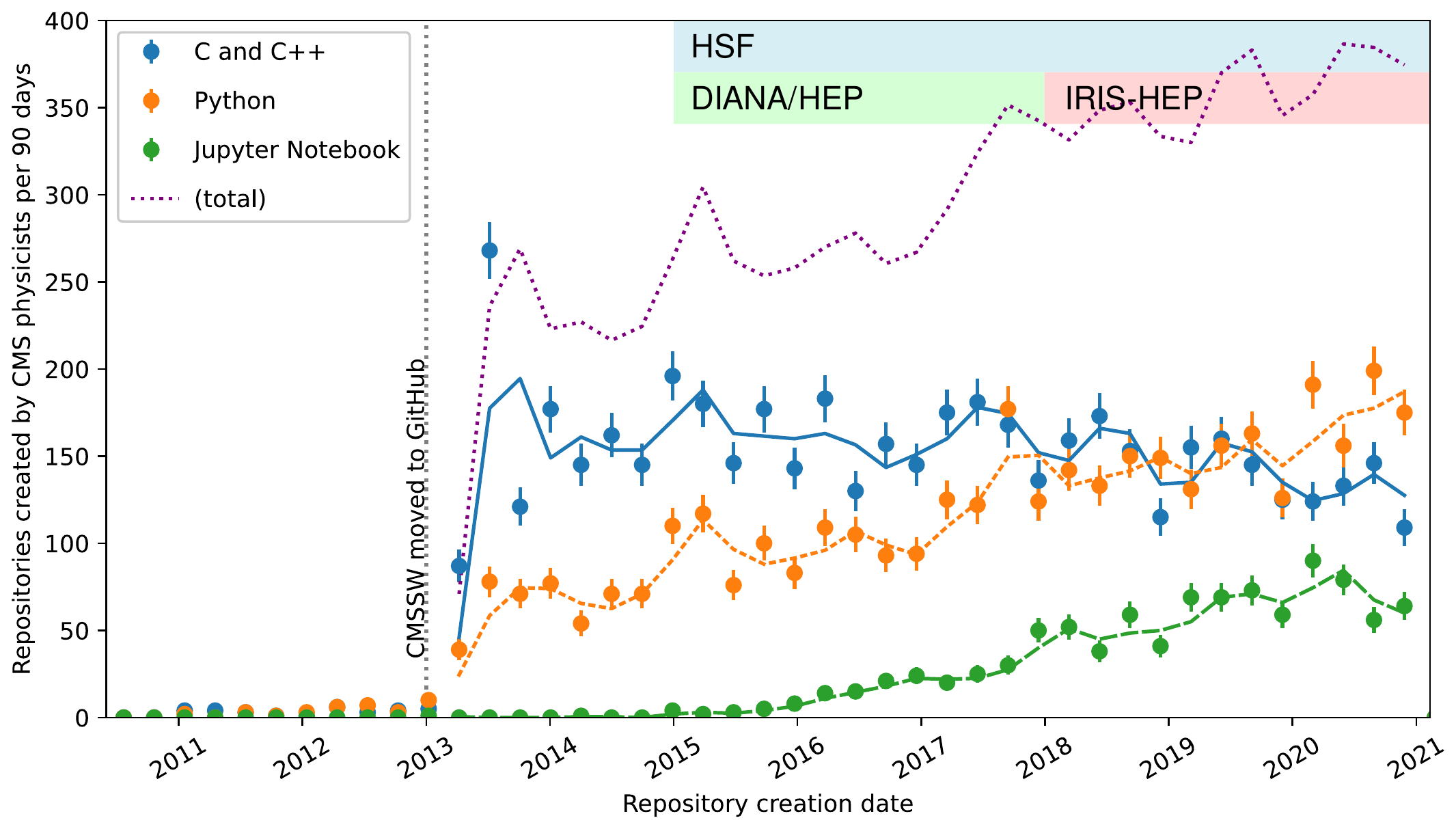}

\caption{Number of GitHub repositories created by CMS physicists by language (exclusive categories), showing the rise of Python and Jupyter. CMS physicists are identified as users who fork {\tt cms-sw/cmssw}. HSF, DIANA/HEP, and IRIS-HEP timelines are overlaid (top-right). \label{fig:lhlhc-github-languages-paper}}
\end{figure}

\begin{figure}
\centering
\includegraphics[width=0.8\linewidth]{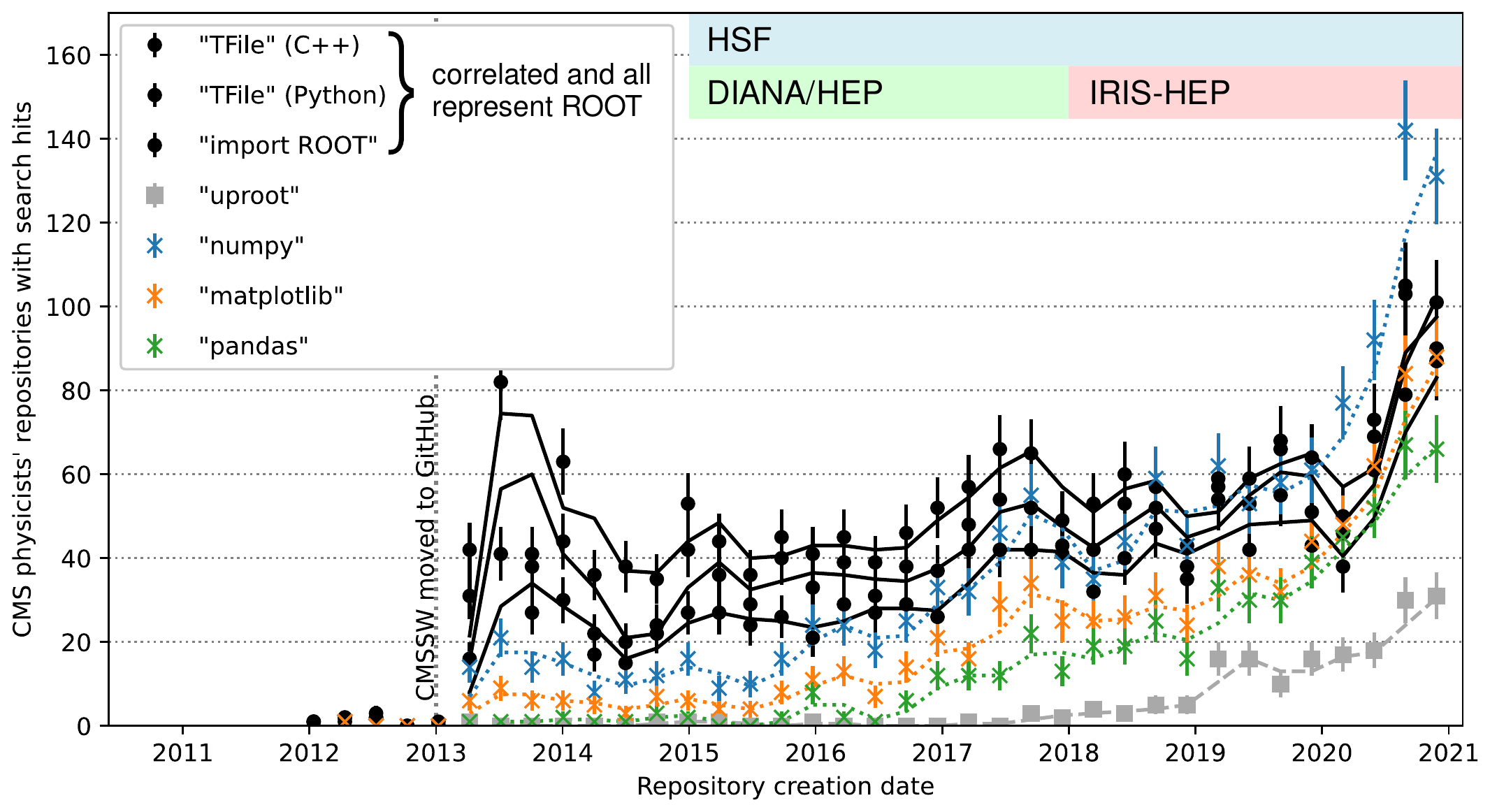}

\caption{Number of GitHub repositories, created by CMS physicists (same analysis as in Figure~\ref{fig:lhlhc-github-languages-paper}), that match search strings (non-exclusive categories). This shows the rise of Pythonic data analysis (NumPy, Matplotlib, Pandas), with ROOT (C++ and Python, also increasing) for scale. \label{fig:lhlhc-github-overlay-lin-paper}}
\end{figure}

Within HEP, adoption of Python also ramped up in the years following 2015. As a specific illustration of this trend, Figure~\ref{fig:lhlhc-github-languages-paper} shows the language of GitHub repositories created by CMS physicists, where CMS physicists are identified as users who forked {\tt cms-sw/cmssw} (\href{https://github.com/jpivarski-talks/2021-02-24-reload-statistics}{link to full analysis}). (A similar analysis of GitLab usage would be limited by GitLab's long-term availability of private repositories.) Python usage in CMS has been steadily growing and surpassed C++ in 2019. GitHub's language assignment is exclusive---each repository is considered entirely Python or entirely C++, but in reality most are mixed---so we can investigate with more precision by counting repositories that match search terms, such as ``numpy,'' ``matplotlib,'' and ``pandas.'' In Figure~\ref{fig:lhlhc-github-overlay-lin-paper}, we see the use of these libraries steadily increasing (long before ``uproot''). Now they appear in repositories as often as ROOT-related terms, such as ``TFile'' and ``import ROOT'' (``from ROOT import'' is less common and highly correlated with the 3 ROOT-related terms shown on the plot).
Similar conclusions can be drawn from an analysis survey run within LHCb in 2018.

\begin{figure}
\centering
\mbox{\hspace{-0.15\linewidth}\includegraphics[width=1.15\linewidth]{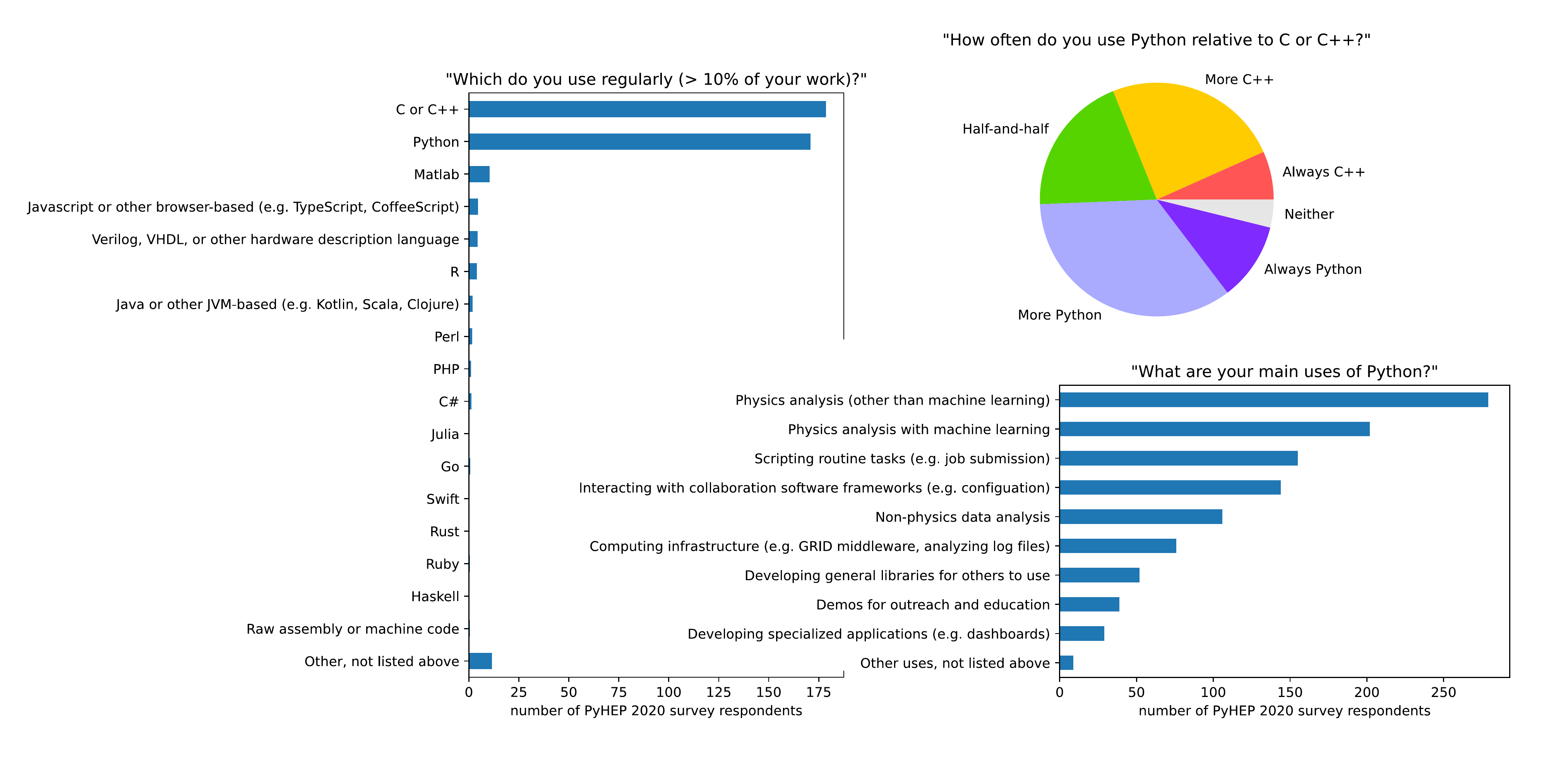}}

\caption{Survey results from 406 registrants of PyHEP 2020~\cite{PyHEP2020} illustrating the mix of C++ and Python (and no other language) among physicists, and the use of Python primarily for analysis. \label{fig:pyhep2020-survey-paper}}
\end{figure}

\begin{figure}
\centering
\includegraphics[width=\linewidth]{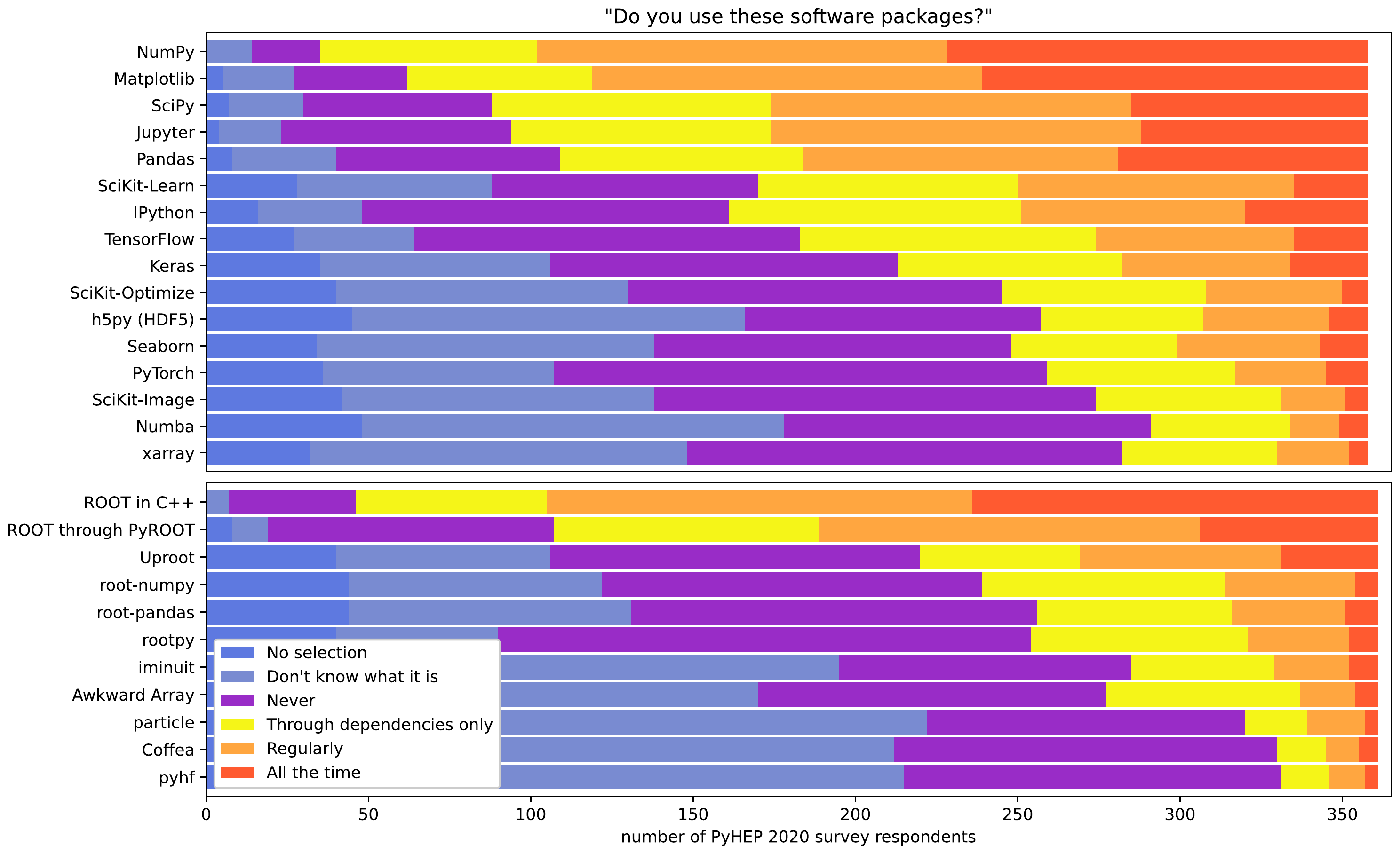}

\caption{Familiarity and usage of data science tools (top) and HEP data analysis tools (bottom) from the same survey as Figure~\ref{fig:pyhep2020-survey-paper}. \label{fig:lhlhc-familiarity-with-packages-paper}}
\end{figure}

A pre-workshop survey sent to the \href{https://indico.cern.ch/e/PyHEP2020}{PyHEP 2020} workshop~\cite{PyHEP2020} registrants also indicated that the HEP community is using an even mix of C++ and Python. Figure~\ref{fig:pyhep2020-survey-paper} shows responses to basic questions and Figure~\ref{fig:lhlhc-familiarity-with-packages-paper} shows the registrants' familiarity and usage of popular data science and HEP tools.

Clearly, HEP analysis culture is in the midst of a transition, perhaps as significant as the transition from Fortran to C++ and object oriented programming in the 1990's~\cite{Brun:2012cq}. The programming languages and software library choices plotted above are the most quantifiable aspects of this transition, but other changes are indicated by the words physicists use to describe their analysis or analysis software:

\begin{itemize}
\item HEP data analysts are becoming averse to monolithic frameworks. ``{\bf Framework}'' (or ``{\bf platform}'') used to be a more positive word: ROOT and collaboration software are self-described as frameworks, and even small analysis groups develop(ed) frameworks for their own ntuple's schemas. A framework, \href{https://root.cern.ch/root/htmldoc/guides/users-guide/ROOTUsersGuide.html#the-framework}{in this sense}, provides all the essentials in one package but requires users to fit their workflows into the application. Data science ``{\bf libraries}'' (or ``{\bf toolkits}''), however, have a very modular mindset, in which narrow scope and interoperability with the rest of the ``{\bf ecosystem}'' are major selling points for a new tool. This mindset is becoming ever more popular in HEP as well, among established tools as well as new ones. New developments in ROOT, for instance, emphasize interoperability~\cite{ROOTTeam:2020jal} and ease of installation~\cite{Shadura_2020} (e.g.\ in conda-forge). It's becoming an expectation that a new library can be added to an existing analysis without disturbing the workflow.

\item Following the Pandas DataFrame model (originally from R), columns---named, typed attributes of all entries in a dataset---are becoming more visible in analysis at the expense of rows, which are instances of identically typed data, usually collision events in HEP. The word ``{\bf columnar analysis}'' is frequently applied to emphasize column-granularity, whether it is for internal data engineering (transferring less data or vectorizing a calculation) or it is highly visible to users as array-at-a-time operations~\cite{Hartmann:2021qzp}. Array-at-a-time interfaces bridge the gap between interactive tinkering (in the style of Pandas or {\tt TTree::Draw}) and production-ready analysis scripts.

\item It has also become common for developers in HEP to describe their tools as ``{\bf declarative}'' (or ``{\bf functional}''). In computer science, a declarative language (like SQL) describes calculations independently of the order in which they would be calculated~\cite{10.5555/895948}. In HEP, this can mean a complete separation of code and data, such as FAST-carpenter~\cite{krikler2020fast}, which specifies cuts and quantities to plot using YAML. It can also refer to domain-specific languages that are as declarative as YAML, such as ADL~\cite{Sekmen:2020vph}, CutLang~\cite{Unel:2021edl}, and func\_adl~\cite{Proffitt:2021wfh}. At the very least, it means a separation of the event loop as a Directed Cyclic Graph (DAG) from the calculations performed on each event, as in ROOT's RDataFrame~\cite{Piparo:2019xdy}, or on a partition of events, as in Coffea's Processor. Since HEP collision events can almost always be processed in any order, a declarative DAG for the event workflow is sufficient to plug HEP analyses into a distributed scheduler, such as Spark, Dask, or Ray.
\end{itemize}

\section{The future of HEP analysis tools}

The previous section describes the past and present of HEP (and particle physics) software. It would be dangerous to extrapolate through the trends: for instance, Figures~\ref{fig:lhlhc-github-languages-paper}--\ref{fig:lhlhc-familiarity-with-packages-paper} indicate that Python and Pythonic data analysis tools (NumPy, Matplotlib, and Pandas) were not widely used in HEP 5~years ago, but now they are as widely used as C++ and ROOT. That does not mean, however, that the trend will continue to rise: there are good reasons to believe that it will level out, with physicists using Python as an interface and C++ for performance and accelerator access (e.g.\ GPUs), freely mixing HEP-specific ROOT routines with machine learning, distributed schedulers, visualization, and other general-purpose tools from the data science world.

This ``mixed future'' is a safe bet because it is how most other scientific fields that adopted Python earlier than HEP operate. AstroPy (2011) is a Pythonic hub for astronomy, but it wraps older C libraries such as WCSLIB, FITS, and HDF5. Geospatial analysis uses a variety of Python tools, but it is largely based on GDAL (2000) and CGAL (1996), both written in C++. Scikit-Learn (2007) and SciPy itself (2001) are primarily wrappers of legacy C and Fortran routines. Unlike Java, Python mixes well with C++, owing to a single, prominent implementation target (CPython) that shares a memory heap with libc/malloc, a well-documented C API, and active C++ binding projects---Cython and pybind11.

\subsection{File formats}

File formats tend to be conservative. The following widely used formats for the sciences are decades old:

\begin{itemize}\setlength{\itemsep}{-0.15 cm}
\item generic: JSON~(2002), XML~(1998), XLS~(1997), HDF5~(1992), NetCDF~(1990), CSV~(1972)
\item Astronomy: VOTable~(2002), FITS~(1981)
\item Geospatial: GeoTIFF~(1994), Shapefiles~(1990)
\item Genetics: FASTA~(1985)
\item Chemistry: ChemML~(1999), GROMACS~(1991), CHARMM~(1983), PDB~(1976)
\end{itemize}
We can therefore expect ROOT files to continue to play a major role in HEP. However, this is only half of the story: interoperability is a key value in the culture of data science software development. With sufficient interoperability, data file formats do not dictate which tools can be used to analyze them. Astronomers, for instance, routinely copy FITS telescope images and VOTable legacy data into machine learning libraries on the fly with NumPy or through on-disk HDF5 files. Geneticists and climate scientists are mixing their old software stacks with xarray (2014), a Pandas-like library for larger-than-memory $N$-dimensional tensors, and Zarr (2015), an array format described as ``cloud-ready NetCDF.''

For many years, ROOT TTrees had been unique in their ability to efficiently store (i.e.\ usually columnar) nested, variable-length data structures with an immediate interpretation in C++~\cite{root-io-survey}. Now, Apache Arrow~\cite{lentner2019shared} (2016) can efficiently represent (always columnar) nested data structures in memory, in a language-independent and interprocess-shareable way. Moreover, Apache Parquet~\cite{vohra2016apache} (2013) can efficiently store (always columnar) these language-independent, nested data structures on disk. As non-domain-specific formats, Arrow and Parquet are recognized across scientific fields and are supported by interdisciplinary libraries like Pandas, Scikit-Learn, and TensorFlow. ALICE's O2 analysis framework~\cite{Alkin:2021mfo} uses Arrow as its primary data model for Run~3.

But that doesn't mean that HEP should ``switch'' from ROOT to Parquet. It means that data analysts will want to mix ROOT, Arrow, Parquet, and HDF5 or even ASDF (2015) in their analysis workflows, and lightweight software needs to be on hand to accommodate that. Uproot is a Python implementation of ROOT I/O, capable of reading and writing TTree data that are sufficiently independent of C++ that they can even have an Arrow/Parquet interpretation~\cite{uproot}. Using Awkward Array, Uproot can avoid creating Python objects for every entry of a TTree during a conversion, which impacts the conversion rate by a factor of hundreds~\cite{Pivarski2017FastAT}. Uproot is also fairly well established among HEP physicists (Figures~\ref{fig:lhlhc-github-overlay-lin-paper} and \ref{fig:lhlhc-familiarity-with-packages-paper}).

It is also important to note that ROOT data will not always be TTrees. The ROOT team is developing a replacement for the TTree class called RNTuple, which addresses TTree's shortcomings relative to Parquet~\cite{Blomer:2020usr}. In particular, RNTuple data are always columnar, little-endian, decouple page boundaries (equivalent to TBasket boundaries) from entry/event boundaries, and have a language-independent interpretation in addition to their C++ interpretation. To achieve the same fluency between RNTuples and Arrow/Parquet as we currently have between TTrees and Arrow/Parquet, either (1)~a Python implementation of the RNTuple format must be developed in the style of Uproot, or (2)~the still-developing RNTuple libraries must maintain their independence from the ROOT codebase to such a degree that a small, RNTuple-only library may be wrapped in Python and distributed on PyPI (with pip). IRIS-HEP's Nebraska team is pursuing option (2), but progress has been slow. If RNTuple becomes a widely used format without a pip-installable library to read and write it, more effort will be needed in this area.

\subsection{Databases}

The most common patterns for data access in physics analysis are actually more suited to databases than files. Typically, experimental collaborations centrally produce analysis datasets with basic physical quantities stored as primitive types. Many analyses will share some subset of these quantities, but each analysis may also require specific variables that are not computed by default. Because of the difficulty and tedium involved in modifying or extending the binary files that compose the centrally-produced datasets, analysts will rerun the entire dataset production just to add their variables in a consistent and usable way. In contrast, a database allows storing and accessing each variable independently, eliminating any duplication of effort, processing, or disk space. A companion product to such a database system is a query language that can succinctly express typical data projections and filters. Commercial data offerings aimed at ``big data’’ have reached similar data scales as HEP; however, they are not yet competitive with HEP domain-specific tools with respect to query language quality or query performance~\cite{Graur:2021isi}.

The Striped project~\cite{Chang:2017ske,Gutsche:2020kmd} implemented a successful prototype of database-style analysis. Along the way, the project developed one of the first non-trivial implementations of columnar analysis in the scientific Python ecosystem, including automatic scaling in batch systems, which grew into Coffea. A mature framework for the usage of databases for analysis has the potential to provide many benefits. Each collaboration could centralize its analysis data in a global, federated, and extensible database, with automatic replication, caching, versioning, provenance tracking, and improved metadata handling. These features, which naturally arise from database-style analysis, would significantly improve both the time to insight and analysis reproducibility. However, because this mode of analysis data management is radically different from what experimental collaborations have pursued to date, it will require a higher level of dedicated effort to put into production.

There are a few projects that are picking up where Striped left off:

\begin{itemize}\setlength{\itemsep}{0 cm}
\item{ServiceX~\cite{Galewsky:2020xig} provides the ability to fill databases directly with derived variables.}
\item{SkyhookDM~\cite{lefevre:login20} is pursuing database storage for analysis data.}
\item{Coffea's columnservice~\cite{columnservice} provides a demonstration of useful configurations and interfaces based on analyst feedback.}
\item{Tiled aims to provide a unified API for data access, which would ease the transition from file-based analysis to broader usage of databases.}
\end{itemize}

Given the many advantages of this approach, described above, increasing effort in this area should be considered high priority.

\subsection{Distributed computing}

The first highly visible data analysis software products from industry addressed the problem of scaling to large datasets, under the catchphrase ``big data.'' A MapReduce framework developed by Google~\cite{62} (2004) was reimplemented in open-source Java as Apache Hadoop~\cite{hadoop} (2006) and was generalized beyond ``map'' and ``reduce'' to a full suite of functional programming primitives in Apache Spark~\cite{zaharia2016apache} (2014). These software products differ from anything used in HEP, other than ROOT's PROOF system~\cite{Ballintijn:2006ni}, in that they mix data analysis logic with scale-out. Traditional batch queue systems run encapsulated code, such as a shell script, that is opaque to the scheduler, so although systems like HTCondor's DAGMan~\cite{htcondor} have map/reduce and task retry features, those features were disconnected from the analysis logic. Big data frameworks can repartition and shuffle intermediate data because the data types in which users write their analyses are a part of the scheduler. 

Apache's decisions to write Hadoop and Spark in Java/Scala made sense because most corporations have a Java-based intranet for business intelligence. It proved difficult for HEP to adopt these frameworks because the Java Virtual Machine (JVM) does not mix well with C++, and therefore any of our tools. Not only are the in-memory data formats different, but Java manages a separate heap from the libc/malloc-based world, freely moving data in RAM (i.e.\ invalidating pointers) as a JVM-dependent detail. Defensively copying data between Java and any Python or C++ process is still slow and brittle.

Meanwhile, Python became the leading data analytics language and dominant machine learning platform (Figure~\ref{fig:analytics-by-language}), and Hadoop/Spark-like tools were developed for Python. Dask~\cite{rocklin2015dask} (2015) takes a highly modular/non-framework approach to this problem: it is primarily a DAG of generic operations, methods to optimize DAGs, and secondarily a scheduler to distribute computations represented by those DAGs. Dask's most popular applications are a distributed array (adhering to NumPy's API) and a distributed DataFrame (adhering to Pandas's API), but the collection types and the distributed processor are built on top of Dask's core---the DAG is the foundation of the library.

In HEP, the Coffea developers experimented with big data scale-out mechanisms, including Striped~\cite{Chang:2017ske,Gutsche:2020kmd} (an in-house Fermilab project), Spark, Parsl, Work Queue, Dask, and Tiled. Of these, Dask has been the most successful so far, with more analyses opting to use the Dask backend than the others. However, this is a case in which fitness for present-day problems might not be sufficient for the HL-LHC: the Dask scheduler is designed for single-user clusters with a single point of failure in the scheduling node, and the largest clusters that have been tested have only involved tens of thousands of nodes.

Fortunately, Dask is modular and other schedulers may be used with it, without interfering with Coffea or physics code. Ray (2017) looks promising: it is poised as a Python-based successor to Apache Spark with significant industry support. Ray lacks a single point of failure, and Dask-Ray integrations are being developed at AnyScale, the company that commercializes Ray.

High-throughput data processing is a generic problem that we can and should solve with open source industry tools. Data delivery is less generic, in that HEP datasets have specialized formats, considerable tooling, and optimizable properties, such as statistically independent events and the columnar layouts of TTrees. Three IRIS-HEP projects, namely ServiceX~\cite{Galewsky:2020xig}, SkyhookDM~\cite{lefevre:login20}, and coffea-casa~\cite{Adamec:2021vkl}, use generic data science tools to build HEP-specific workflows. These are good examples of the ``mixed future,'' in which Docker Kubernetes, Helm, Minio, Flask, RabbitMQ, Kafka, Ceph, and Gandiva are used alongside ROOT, Rucio, XCache, and Uproot to deliver columns of data to analyses as Arrow or Awkward Array buffers, Parquet or ROOT files. The data science components and the HEP components can fit together because they share values of modularity, sharply defined interfaces, and columnar data granularity.

\subsection{Acceleration}

It may be surprising that Python emerged as the language of choice for large-scale data analysis, especially machine learning, when it has so many language features that prevent fast computation: runtime type checking, garbage collection, boxing numbers as objects, no value types or move semantics at all (all Python references are ``pointer chasing''), virtual machine indirection, and a Global Interpreter Lock (GIL) that prevents threads from running in parallel. Some features are particular to the CPython implementation, but this is by far the most commonly used: most extensions written in C or C++ don't work with alternatives. Python is popular for performance-demanding applications despite these handicaps because developers have learned to split programs into a fast, simple part (in a compiled extension) and a slow, complex part (in pure Python). Python's dynamic features make it easier to deal with complexity, and the number-crunching can always be extracted from Python into a compiled extension, or even expressed in primitives provided by existing compiled extensions, such as NumPy.

The separation into fast-simple versus slow-complex is a good one to make even if not forced to do so by the language. Legacy C++ applications are often difficult to port to GPUs because their number-crunching logic is mixed with ``bookkeeping,'' for instance as a suite of interacting class instances, which can prevent SIMD-friendly refactoring.

Software developers can write compiled extensions, but data analysts need lower barriers to optimizing their code---not every script justifies a new Python extension module. Numba~\cite{lam2015numba} (2012) provides the lowest barrier to compilation, since it Just-In-Time (JIT) compiles Python code for CPU and GPU backends. The effect is immediate: prepending a function definition with ``{\tt @numba.jit}'' compiles that function so that it runs up to hundreds of times faster. Many HEP packages use Numba to accelerate Python code, including Scikit-HEP's Awkward Array, Hist, and Vector, as do some collaboration frameworks, such as Xenon1T's trigger pipeline. Numba's drawback is that it implements a subset of Python to allow static typing, and it can be confusing when code that works in Python doesn't compile in Numba.

ROOT has offered JIT-compilation of C++ for a long time as Cling~\cite{Vasilev_2012} (replacing its former CINT interpreter), as well as dynamic Python/C++ bindings as PyROOT/Cppyy~\cite{Lavrijsen:2015khc}. Just as Numba allows physicists to JIT-compile a Python function, ROOT JIT-compiles a string of C++ code with automatic bindings. On the one hand, physicists must understand two languages, but on the other, this C++ is not a restricted subset, but the full C++ language, including CUDA for GPUs and any hardware accelerator with a C++ interface. RDataFrame is a particularly good interface for organizing and distributing tasks written in C++ (on Spark and Dask).

However, Cling's JIT-compilation facility is currently maintained as part of the ROOT project, when it should be upstreamed to LLVM as part of Clang. Princeton's Compiler-as-a-Service (CaaS) project~\cite{compiler_research} is working on integrating Cling's LLVM patches into the main LLVM project, where they can be maintained by the larger community. The same project is also making PyROOT/Cppyy's dynamic language bindings, currently unique to HEP, into a standalone project that can be adopted by data scientists beyond HEP.

In principle, the Julia language~\cite{bezanson2017julia} would be ideal for HEP, as it combines Python's dynamism with C++'s speed by JIT-compiling every expression in the language. Whereas Numba can only JIT-compile a subset of Python and both Python and C++ struggle with adding autodifferentiation (using JAX in Python and CaaS's Clad in C++), this is a built-in feature of Julia. Julia lacks the very large world-wide community of both Python and C++, though Python and C++ interoperability is very active in the Julia community. Small groups of physicists have been investigating Julia for years---it could develop into a third HEP language on the HL-LHC's timescale.

\subsection{Histogramming}

Whereas the previous two sections, distributed computing and acceleration (horizontal and vertical scaling), are general concerns beyond HEP and should be addressed by non-HEP software products, histogramming is more HEP-specific. The statistical concept of approximating a distribution with binned counts is generic, but the way that physicists use histograms as fillable, mergeable objects, is unique. Most plotting libraries provide a histogram function to {\it plot} data as a binned histogram, but HEP users expect to be able to fill histogram objects, save them to disk, add them, weight them e.g. by relative luminosities, and fit them with parameterized curves or more general statistical methods. All of the software products we know of with these features have some connection to HEP---developed for HEP or by a physicist who has left the field.

Histogram additivity is useful because we are interested in distributions of enormous numbers of statistically independent events, which isn't true of all fields of data analysis. This paradigm is general enough to think that our histogram-centric way of doing analysis might catch on beyond HEP, but only if our tools can be easily installed and used in the wider ecosystem.

Until recently, histogramming in Python was too fractured to be productive. There were at least 21 pip-installable packages for histogramming on PyPI before April 2019, not counting major projects like ROOT and YODA (not pip-installable). None had wide adoption within the community. All of these implement basic functionality, some with fast filling performance, but none were a ``complete'' package for HEP analysis. The most complete, Physt, couldn't handle negative weights, for instance. It is not difficult to create a basic histogramming library in Python, but it is difficult to make it general enough for all use-cases.

The inclusion of a C++ histogram library into Boost (by a HEP physicist) and a Python wrapper for this library in late 2019 within the Scikit-HEP project changed the situation by providing a ``nucleation site'' for histogram development. As a focused, single-purpose high-performance library, Boost-histogram~\cite{henry_schreiner-proc-scipy-2020} fits the data science ideal of modularity, and its association with Boost ensured a longevity that incentivized building on top of this library, rather than competing with it by building new libraries from scratch. As a case in point, the Hist~\cite{Dembinski:2020dic} library adds analyst-friendly features (such as plotting) that the no-frills Boost-histogram lacks, but it does so as a subclass of the Boost-histogram type. Two libraries with the same functionality can be complementary if one uses the other to provide a different level of abstraction.

Beyond packaging, the increasing use of template fits in HEP analysis is changing how histograms are used. Traditionally, histograms have represented data that a human might inspect and fit with a parameterized curve. Template fits use the histograms themselves as the curve and tune weights required to fit one weighted sum of histograms to another. Data in many signal and background samples must be subdivided into signal, control, and validation regions, and they must be filled under thousands of systematic variations. The data for these fits may be thought of as weighted event counts in voxels of a many-dimensional space, usually more than the 3 dimensions that a human can inspect, and some of these dimensions are not continuous, such as ``number of $b$-tags.'' Binning dimensions for systematic variations represent a function, rather than a distribution, which differs from the original purpose of histogramming, but is well-defined if applied consistently.

Modern histogramming libraries, such as Boost-histogram and ROOT's v7 histograms, allow arbitrarily many dimensions with different dimension types (categorical and continuous), but they do not manage filling a distribution along some dimensions and functional variations along others. Special tools like HistFitter~\cite{Besjes:2015vns}, TRexFitter, cabinetry~\cite{cranmer_kyle_2021_4627038}, and FAST-carpenter fill this niche with a higher level of abstraction above histograms. Declarative configurations (usually YAML) describe how to subdivide the data into regions, how to bin axes, and how to vary systematics. Template fitters like Combine~\cite{CMS-NOTE-2011-005}, HistFactory~\cite{Cranmer:2012sba}, and pyhf~\cite{pyhf_joss} perform the fits, make plots, and set statistical limits.

While these developments are improving the statistical reach of LHC analyses, the histograms that comprise the template data are not often serialized efficiently. Thousands of histograms in a template-set have the same or mostly the same axes: the usual approach of saving them as separate histogram objects in a ROOT file duplicates the axis metadata in independently compressed buffers. The common structure of template-sets would benefit from a columnar approach, in which all or large chunks of the numeric data are stored in contiguous buffers and shared metadata (axes, labels, etc.) are stored exactly once. HDF5 and Zarr are well-suited to efficiently store and retrieve a many-dimensional array like this.

In addition to efficient serialization, histogram formats (Boost Histogram, ROOT, YODA, HEPData YAML) and views, such as plottable histograms, components of a template-set, raw arrays, and tables such as Pandas and xarray, need to be more interconvertible. Each of these formats and views have distinct benefits. Being able to fluidly convert among them all would have a multiplying effect on analysis productivity.

\subsection{Fitting and statistics}

Software that performs fits and reveals statistical limits or measurements with uncertainties needs to be even more customized for HEP use-cases than histogramming. Even the traditional case of fitting to a parameterized curve is not entirely common beyond HEP, since first-principles or approximate functional forms are more often achievable in HEP and related fields than those that study human behavior (e.g.\ targeted advertising). Today, HEP analysts engage in much more refined---and specialized---fits than traditional curve-fitting, and the development of new statistical techniques is often viewed as a part of the process of doing analysis in HEP.

As such, there are a lot of small, focused packages in this area. The most foundational is iminuit~\cite{iminuit} (2015), following the now-deprecated pyminuit (2008), which provides Minuit~\cite{James:1975dr} minimization in Python so that users can build binned and unbinned fits from the ground up. Scikit-HEP's iminuit is a good example of the ``mixed future,'' in which a decades-old algorithm developed in HEP (Minuit, 1975) is fully integrated into NumPy, SciPy, Jupyter, Numba, JAX, and Matplotlib. iminuit has a larger userbase among astronomers than physicists because astronomers adopted Python earlier than HEP. Packages like probfit (2012, now deprecated) and zfit~\cite{Eschle:2020ghu} (2019) provide high-level curve-fitting interfaces, with zfit leveraging TensorFlow and using a variety of minimizers as backends: iminuit, nlopt, TensorFlow, and SciPy. Whereas Minuit focuses on detailed few-parameter fits (fewer than 100), optimizers for deep learning like TensorFlow can optimize thousands of parameters at once, but provide less detail about the shape of the minimum for uncertainty bands.

For template binned fits, Scikit-HEP's pyhf (2018) is a major project, also with a variety of backends: NumPy, TensorFlow, PyTorch, and JAX. Two of these, TensorFlow (through an explicit DAG) and JAX (through autodifferentiation) differentiate the objective function for faster convergence, and all but NumPy can take advantage of GPUs. pyhf is implemented in Python without ROOT dependencies, but is typically much faster than ROOT-based template fitters, such as HistFitter, because it is leveraging exact derivatives and hardware acceleration through its backends. For example, one fully optimized model took 10~hours to compute in HistFitter and 30~minutes in pyhf, and another took 20~minutes in HistFitter and 10~seconds in pyhf. pyhf is (so far) cited in 19~publications and is regularly used to derive limits in ATLAS, 18~of which are published as full likelihood functions~\cite{ATL-PHYS-PUB-2019-029} in the HEPData repository, which allows the experimental result to be applied to new theoretical models in the future.

Statistical inference is such a broad field that there is a long tail of specialized packages. To get a sense of the variety, here are a few from PyPI:

\begin{itemize}
\item Visualization with statistical interpretation: quickstats, heppi, hep\_spt, scikit-validate (comparison with baseline), ClusterKinG (cluster analysis), hepstats (Bayesian Blocks for histogram optimization, hypothesis tests, splot generators)
\item Steps in an analysis: PyOD (outlier detection), raredecay (reweighting and classification), PynFold (unfolding), selanneal (optimizing cuts), pyBumpHunter (bump hunting), simplify (simplifying likelihood functions), fastjet (Python wrapper of the FastJet library)
\item HEP-specific interfaces to data science tools: MadMiner (machine learning inference), Hepynet (deep neural networks)
\item Physics-specific generators and optimizers: dEFT-HEP (fitting Effective Field Theories), smelli (likelihoods for Standard Model Effective Field Theory), wilson (running Standard Model Effective Field Theories), minimal-lagrangians (for dark matter models), mt2 (stransverse mass variable), Hazma (dark matter limits), PhaseSpace (phase space generation using TensorFlow).
\end{itemize}

The small, specialized package approach can be contrasted with distributing statistical tools within a framework like ROOT. The advantage of creating a new package and putting it on PyPI, which makes it pip-installable by anyone with a Python distribution, is that the developer completely controls the timescale in which updates can be distributed to users. This is particularly useful for statistical techniques that develop rapidly, and for correcting bugs discovered by users. The same procedure, distributed as a class in ROOT, would involve a pull review/merging process before it can be accepted into the main codebase and would only reach users (other than hand-picked testers) on ROOT's deployment timescale, which is once every 1--5~months and can also depend on constraints imposed by the LHC experiments. Moreover, if the original developers abandon a Python package, it can simply be ignored, but an abandoned class in ROOT has to be maintained indefinitely by the ROOT developers.

On the other hand, independent packaging puts the onus of good packaging practices on the developers, who would often rather focus on physics or statistics. Good packaging is more difficult than it at first appears: it's easy to assume (explicitly or implicitly) that the user's computer is set up like the developer's, often through unlisted dependencies, leading to installation problems. Modern packaging tools guard against this, but outdated instructions that don't use the new tools are easy to find online, making it hard for new developers to know what to do.

Scikit-HEP, originally conceived as a metapackage for HEP, evolved quickly towards a community-oriented and community-driven project providing a collection of well-scoped libraries that form a "Big Data" analysis ecosystem for HEP (together with common libraries beyond HEP). As a GitHub organization, Scikit-HEP adopts well-maintained repositories contributed by independent developers so that they are mutually discoverable, and it enforces high standards for packaging, testing, documentation and software architecture. To promote quality packaging, the \href{https://github.com/scikit-hep/cookie}{scikit-hep/cookie} package provides a recipe for cookiecutter, which uses command-line questions to generate a package skeleton with the current best practices: declarative configuration (setup.cfg, pyproject.toml, MANIFEST.in), automatic version-generation (setuptools\_scm), compiler tools for extensions (pybind11 and cibuildwheel), simplified environments for pure Python packages (filt and poetry), automated testing (pytest and GitHub Actions), dependency updates (Dependabot), formatting and linting (pre-commit, MyPy, Flake8, Black, isort, PyUpgrade), and documentation stubs (Sphinx and ReadTheDocs). It is much easier to use a generated template, even if an existing package has to be ported into it, than it is to follow a list of rules.

What this does not solve, however, is the problem of discovery. Physicists often create packages without knowing that they duplicate functionality, or without reaching potential users, or without knowing packaging best practices that would solve many common installation issues. This is a general non-HEP-specific problem of building a community: with a well-defined community, a physicist who wants to share a statistical technique as a package would have heard of similar projects and would be aware of where to look for guidance. The HSF PyHEP Working Group and its events have increasingly been filling that role, with hundreds of attendees at the 2020 and 2021 workshops, Module of the Month demos and active Gitter channels. IRIS-HEP is also an intellectual hub for analysis tool development, mainly through its very active Slack workspace. However, the nature of analysis tool development is very distributed, and these support networks haven't permeated through the entire HEP community.

\subsection{Relationship to the community outside of HEP}

As described above, we don't expect all of our software needs to be satisfied by the data science industry, nor should all the software we use be written by physicists. In general, software functionality fits into the following categories, with a spectrum between them:

\begin{enumerate}
\item The functionality in question is not at all HEP-specific and is well-developed outside of HEP, such as machine learning foundations, distributed computing, JIT-compilation, autodifferentiation, and computing interfaces such as notebooks and other code editors.

\item The functionality is in principle not HEP-specific, but also not well-developed outside of HEP. Examples include Cling (the C++ JIT-compiler), Awkward Array (for manipulation of irregularly shaped data), histograms-as-objects, and some statistical techniques like template fits and unfolding.

\item The functionality is and will likely always be HEP-specific, such as tools for manipulating Lorentz vectors (\href{https://github.com/scikit-hep/vector}{Vector}) or for applying energy corrections (\href{https://github.com/cms-nanoAOD/correctionlib}{correctionlib}), tools dealing with particle information (Particle) and decay files and chains (\href{https://github.com/scikit-hep/decaylanguage}{DecayLanguage}).
\end{enumerate}

Given the upcoming computational challenges of the HL-LHC and the maturity of data science software today, it would be prudent to adopt data science solutions in category (1), and to contribute back or proselytize HEP-developed solutions in category (2), so that they can be maintained by the larger community as well. In category (2), a project like Cling has the LLVM community to receive it, once technical issues are overcome, but a project like Awkward Array or Hist would have to convince data scientists beyond HEP that non-tabular data or histogram objects are useful concepts before we can expect collaboration. Better interoperability between HEP tools and data science tools aids sharing in both directions: generic tools become usable in HEP (so that we do not need to maintain them) and HEP-style techniques may become common beyond HEP (so that we do not need to maintain them). In both directions, the key ingredient is interoperability.

\section{Management, risk assessment, and the Grand Challenges}

The use of data science software in HEP is not centrally managed, but HSF/PyHEP, IRIS-HEP, and major community projects such as Scikit-HEP and Coffea, and package developers, have a strong influence on its direction and priorities. IRIS-HEP, in particular, is funding the development of core pieces of the ecosystem.
Development goals are incremental and adjust to what developers of related packages are working on. Through constant communication in the community HSF and PyHEP Gitter channels, as well as other, smaller-scoped channels and the IRIS-HEP Slack project, these teams have a shared idea of what data science/HEP interoperability would and should look like, and are collaboratively building up that ecosystem. Different teams and individuals may have slightly different goals---for instance, Coffea users need to analyze present-day (CMS) data and IRIS-HEP is focused on the HL-LHC---but these kinds of differences can be accommodated. For example, Coffea needed histogram and Lorentz vector objects on a shorter timescale than IRIS-HEP developers could produce a future-proof foundation, so both teams agreed for Coffea to develop ``quick and dirty'' components that were later replaced by the IRIS-HEP ones. Coffea's implementation of both served as prototypes that informed the IRIS-HEP development.

Communication through PyHEP Gitter, IRIS-HEP Slack, and issues/pull requests in GitHub is more frequent and a denser network between teams than the conventional HEP style of slide-presentations in weekly meetings. However, the problem of discoverability is as prevalent as ever: software developers have to ``find'' this network to get oriented. It is still the case that physicist-developers build packages in isolation, only to find out that they could have multiplied their effectiveness if they had been plugged into this ecosystem earlier.

Progress is generally measured in terms of physics analysis productivity in current LHC analyses, which is not the same thing as raw computational performance. Given the way that development self-adjusts to present conditions, the risk is that scaling to the HL-LHC will be ignored. For this reason, IRIS-HEP is conducting an institute-wide ``Grand Challenge'' as an integration test of all its components~\cite{grand_challenge}.
The data processing component of this challenge, conducted with the U.S. LHC Operations program, the ATLAS and CMS collaborations, and the WLCG, is a series of global processing tasks in 2021, 2023, 2025, and 2027 that slowly scale up to the throughput and features expected of HL-LHC data processing: an exabyte of raw data in 100~days.

\begin{figure}
\centering
\includegraphics[width=\linewidth]{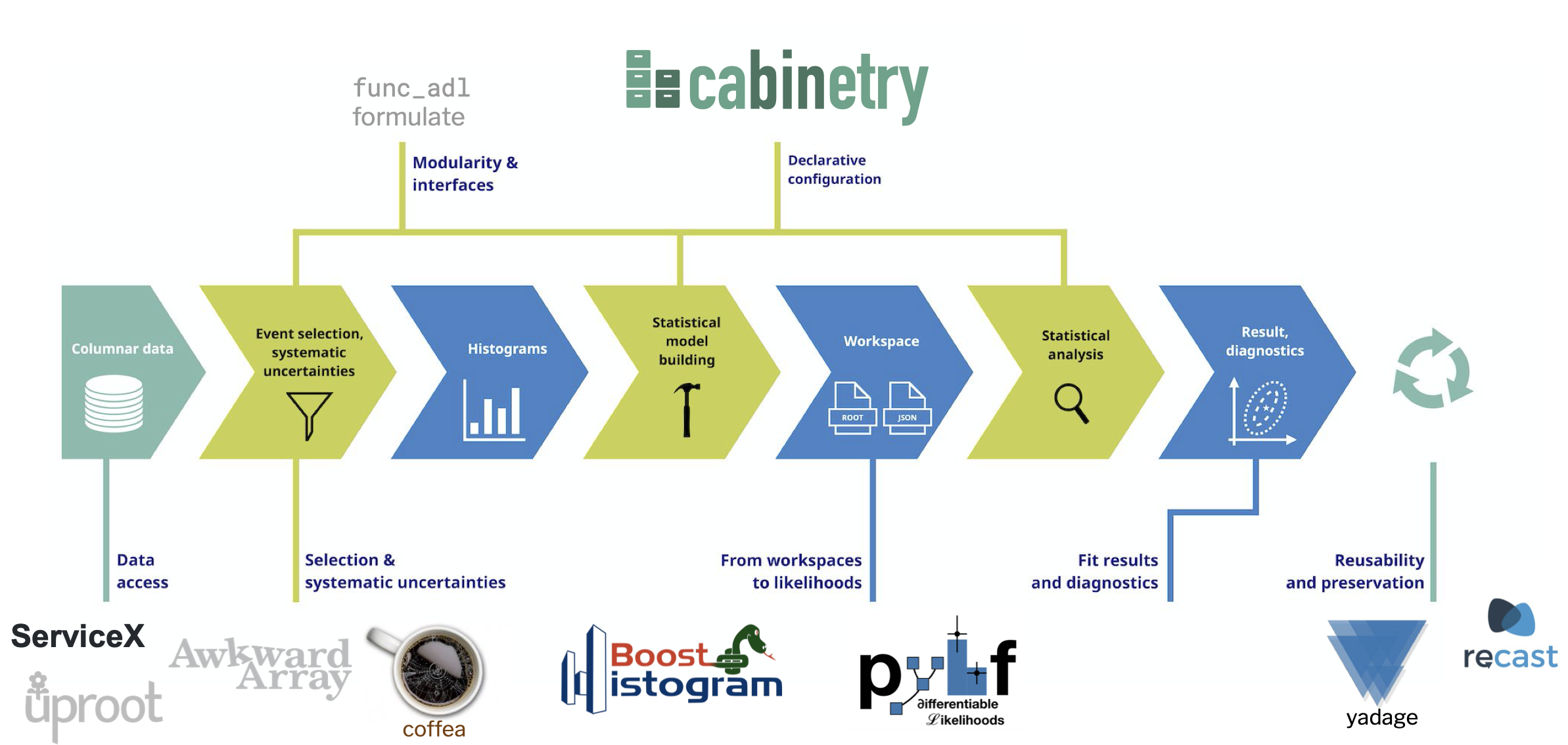}

\caption{Analysis system workflow for IRIS-HEP's Analysis Grand Challenge: how the software libraries fit into an ``end user'' analysis, from analysis-ready data to final results and reinterpretations. \label{fig:cabinetry-vertical-slice}}
\end{figure}

The analysis component of this challenge stresses both computational performance and tool integration. Analysis systems start with the output of the data processing step, about 200~TB of analysis-ready data per physics study, with many studies requesting overlapping subsets of the data. Perhaps more importantly, an analysis workflow integrates many of the services and software libraries described above, so the Analysis Grand Challenge will focus on interoperability. The Challenge will reproduce a standard analysis, such as the $H \to 4\ell$ discovery, with all corrections and systematics studies performed by the Python/data science toolchain, presented in Figure~\ref{fig:cabinetry-vertical-slice}. In addition, the Analysis Grand Challenge will include autodifferentiation as a test of novel techniques.

By testing all of the services and software libraries in a combined workflow, the IRIS-HEP Grand Challenges demonstrate that no major capabilities are missing, verify that each component can efficiently hand off data to the next, and determine if they can, indeed, perform at HL-LHC scales. Starting these challenges now ensures that the developer community has enough time to respond to any issues that it reveals.

\section{Bibliography}

\printbibliography

@misc{HSF,
title = {{HEP Software Foundation (HSF)}},
url = "https://hepsoftwarefoundation.org/",
}

@misc{DIANA/HEP,
title = {{Data-Intensive ANAlysis for High Energy Physics (DIANA/HEP)}},
url = "http://diana-hep.org/",
}

@misc{IRIS-HEP,
title = {{Institute for Research and Innovation in Software for High Energy Physics  (IRIS-HEP)}},
url = "https://iris-hep.org/",
}

@misc{hsf-workshop-slac2015,
author={{Boehnlein}, Amber and {Mato}, Pere and {Mount}, Richard and {Wenaus}, Torre},
title={{The HEP Software Foundation Workshop}},
month={Jan},
year={2015},
url = "https://indico.cern.ch/event/357737",
}

@misc{PyHEP2020,
title = {{PyHEP 2020 Workshop}},
url = "https://indico.cern.ch/e/PyHEP2020",
}

@article{Albrecht2019,
author={HEP Software Foundation},
title="{A Roadmap for HEP Software and Computing R{\&}D for the 2020s}",
journal={Computing and Software for Big Science},
year={2019},
month={Mar},
day={20},
volume={3},
number={1},
pages={7},
issn={2510-2044},
doi={10.1007/s41781-018-0018-8},
url={https://doi.org/10.1007/s41781-018-0018-8}
}

@article{Shadura_2020,
	doi = {10.1088/1742-6596/1525/1/012050},
	url = {https://doi.org/10.1088/1742-6596/1525/1/012050},
	year = 2020,
	month = {apr},
	publisher = {{IOP} Publishing},
	volume = {1525},
	pages = {012050},
	author = {O Shadura and B Bockelman and V Vassilev},
	title = {Evolution of {ROOT} package management},
	journal = {Journal of Physics: Conference Series}
}

@article{Hartmann:2021qzp,
    author = {Hartmann, Nikolai and Elmsheuser, Johannes and Duckeck, G\"unter},
    collaboration = "ATLAS Software and Computing",
    title = "{Columnar data analysis with ATLAS analysis formats}",
    doi = "10.1051/epjconf/202125103001",
    journal = "EPJ Web Conf.",
    volume = "251",
    pages = "03001",
    year = "2021"
}

@article{Pivarski2017FastAT,
  title={Fast access to columnar, hierarchically nested data via code transformation},
  author={J. Pivarski and P. Elmer and B. Bockelman and Zhe Zhang},
  journal={2017 IEEE International Conference on Big Data (Big Data)},
  year={2017},
  pages={253-262}
}

@article{Pivarski:2020txo,
    author = "Pivarski, Jim and Lange, David and Elmer, Peter",
    title = "{Nested data structures in array frameworks}",
    doi = "10.1088/1742-6596/1525/1/012053",
    journal = "J. Phys. Conf. Ser.",
    volume = "1525",
    number = "1",
    pages = "012053",
    year = "2020"
}

@techreport{10.5555/895948,
author = {Lloyd, J. W.},
title = {{Declarative Programming in Escher}},
year = {1995},
publisher = {University of Bristol},
address = {GBR}
}

@inproceedings{krikler2020fast,
  title={{The FAST-HEP toolset: Using YAML to make tables out of trees}},
  author={Krikler, Benjamin Edward and Davignon, Olivier and Kreczko, Lukasz and Linacre, Jacob},
  booktitle={EPJ Web of Conferences},
  volume={245},
  pages={06016},
  year={2020},
  organization={EDP Sciences}
}

@article{Sekmen:2020vph,
    author = "Sekmen, Sezen and Gras, Philippe and Gray, Lindsey and Krikler, Benjamin and Pivarski, Jim and Prosper, Harrison B. and Rizzi, Andrea and Unel, Gokhan and Watts, Gordon",
    editor = "Mansoulie, Bruno and Marchiori, Giovanni and Salern, Roberto and Bos, Tulika",
    title = "{Analysis Description Languages for the LHC}",
    eprint = "2011.01950",
    archivePrefix = "arXiv",
    primaryClass = "hep-ph",
    reportNumber = "FERMILAB-PUB-21-145-SCD",
    doi = "10.22323/1.382.0065",
    journal = "PoS",
    volume = "LHCP2020",
    pages = "065",
    year = "2021"
}

@article{Unel:2021edl,
    author = "Unel, G. and Sekmen, S. and Toon, A. M. and Gokturk, B. and Orgen, B. and Paul, A. and Ravel, N. and Setpal, J.",
    title = "{CutLang V2: towards a unified Analysis Description Language}",
    eprint = "2101.09031",
    archivePrefix = "arXiv",
    primaryClass = "hep-ph",
    doi = "10.3389/fdata.2021.659986",
    month = "1",
    year = "2021"
}

@article{Proffitt:2021wfh,
    author = "Proffitt, Mason and Watts, Gordon",
    title = "{FuncADL: Functional Analysis Description Language}",
    eprint = "2103.02432",
    archivePrefix = "arXiv",
    primaryClass = "physics.data-an",
    doi = "10.1051/epjconf/202125103068",
    journal = "EPJ Web Conf.",
    volume = "251",
    pages = "03068",
    year = "2021"
}

@article{Piparo:2019xdy,
    author = "Piparo, Danilo and Canal, Philippe and Guiraud, Enrico and Valls Pla, Xavier and Ganis, Gerardo and Amadio, Guilherme and Naumann, Axel and Tejedor, Enric",
    editor = "Forti, A. and Betev, L. and Litmaath, M. and Smirnova, O. and Hristov, P.",
    title = "{RDataFrame: Easy Parallel ROOT Analysis at 100 Threads}",
    reportNumber = "FERMILAB-CONF-19-550-SCD",
    doi = "10.1051/epjconf/201921406029",
    journal = "EPJ Web Conf.",
    volume = "214",
    pages = "06029",
    year = "2019"
}

@misc{root-io-survey,
  author="Jim Pivarski",
  title="Survey of columnar file formats and the techniques they use",
  month="Feb",
  year="2017",
  url = "https://indico.fnal.gov/event/13665/\#9-survey-of-columnar-file-form",
}

@incollection{lentner2019shared,
  title={Shared Memory High Throughput Computing with {Apache Arrow™}},
  author={Lentner, Geoffrey},
  booktitle={Proceedings of the Practice and Experience in Advanced Research Computing on Rise of the Machines (learning)},
  pages={1--2},
  year={2019}
}

@incollection{vohra2016apache,
  title={{Apache Parquet}},
  author={Vohra, Deepak},
  booktitle={Practical Hadoop Ecosystem},
  pages={325--335},
  year={2016},
  publisher={Springer}
}

@article{Alkin:2021mfo,
    author = "Alkin, Anton and Eulisse, Giulio and Grosse-Oetringhaus, Jan Fiete and Hristov, Peter and Kabus, Maja",
    title = "{ALICE Run 3 Analysis Framework}",
    doi = "10.1051/epjconf/202125103063",
    journal = "EPJ Web Conf.",
    volume = "251",
    pages = "03063",
    year = "2021"
}

@misc{uproot,
  author="Jim Pivarski",
  title={{Uproot: rapidly moving data from ROOT to Numpy and Pandas}},
  month="Feb",
  year="2018",
  url = "https://indico.cern.ch/event/686641/\#4-uproot-rapidly-moving-data-f",
}

@article{Blomer:2020usr,
    author = "Blomer, Jakob and Canal, Philippe and Naumann, Axel and Piparo, Danilo",
    editor = "Doglioni, C. and Kim, D. and Stewart, G. A. and Silvestris, L. and Jackson, P. and Kamleh, W.",
    title = "{Evolution of the ROOT Tree I/O}",
    eprint = "2003.07669",
    archivePrefix = "arXiv",
    primaryClass = "cs.DB",
    reportNumber = "FERMILAB-CONF-20-165-SCD",
    doi = "10.1051/epjconf/202024502030",
    journal = "EPJ Web Conf.",
    volume = "245",
    pages = "02030",
    year = "2020"
}

@article{ROOTTeam:2020jal,
    author = "Amadio, Guilherme and others",
    collaboration = "ROOT Team",
    title = "{Software Challenges For HL-LHC Data Analysis}",
    eprint = "2004.07675",
    archivePrefix = "arXiv",
    primaryClass = "physics.data-an",
    month = "5",
    year = "2020"
}

@article{Brun:2012cq,
    author = "Brun, Ren",
    editor = {Ernst, Michael and D\"ullmann, Dirk and Rind, Ofer and Wong, Tony},
    title = {{The Evolution of Software in High Energy Physics}},
    doi = "10.1088/1742-6596/396/5/052016",
    journal = "J. Phys. Conf. Ser.",
    volume = "396",
    pages = "052016",
    year = "2012"
}

@article{Chang:2017ske,
    author = "Chang, Jin and Gutsche, Oliver and Mandrichenko, Igor and Pivarski, James",
    title = "{Striped Data Server for Scalable Parallel Data Analysis}",
    reportNumber = "FERMILAB-CONF-18-016-CD",
    doi = "10.1088/1742-6596/1085/4/042035",
    journal = "J. Phys. Conf. Ser.",
    volume = "1085",
    number = "4",
    pages = "042035",
    year = "2018"
}

@article{Gutsche:2020kmd,
    author = "Gutsche, Oliver and Mandrichenko, Igor",
    editor = "Doglioni, C. and Kim, D. and Stewart, G. A. and Silvestris, L. and Jackson, P. and Kamleh, W.",
    title = "{Striped Data Analysis Framework}",
    reportNumber = "FERMILAB-CONF-20-632-LDRD-SCD",
    doi = "10.1051/epjconf/202024506042",
    journal = "EPJ Web Conf.",
    volume = "245",
    pages = "06042",
    year = "2020"
}

@misc{columnservice,
author="{Nicholas Smith and Burt Holzman}",
title="{ColumnService: a multi-tenant service for caching columnar data}",
url = "https://github.com/CoffeaTeam/columnservice",
}

@article{Smith:2020pxs,
    author = "Smith, Nicholas and others",
    editor = "Doglioni, C. and Kim, D. and Stewart, G. A. and Silvestris, L. and Jackson, P. and Kamleh, W.",
    title = "{Coffea: Columnar Object Framework For Effective Analysis}",
    eprint = "2008.12712",
    archivePrefix = "arXiv",
    primaryClass = "cs.DC",
    reportNumber = "FERMILAB-CONF-20-494-CMS-SCD",
    doi = "10.1051/epjconf/202024506012",
    journal = "EPJ Web Conf.",
    volume = "245",
    pages = "06012",
    year = "2020"
}

@article{Graur:2021isi,
    author = {Graur, Dan and M\"uller, Ingo and Proffitt, Mason and Fourny, Ghislain and Watts, Gordon T. and Alonso, Gustavo},
    title = "{Evaluating Query Languages and Systems for High-Energy Physics Data}",
    eprint = "2104.12615",
    archivePrefix = "arXiv",
    primaryClass = "cs.DB",
    month = "4",
    year = "2021"
}

@inproceedings{62,
title	= {{MapReduce:} Simplified Data Processing on Large Clusters},
author	= {Jeffrey Dean and Sanjay Ghemawat},
year	= {2004},
booktitle	= {OSDI'04: Sixth Symposium on Operating System Design and Implementation},
pages	= {137--150},
address	= {San Francisco, CA}
}

@misc{hadoop,
  author = {{Apache Software Foundation}},
  title = {Hadoop},
  url = "https://hadoop.apache.org",
  day = "19",
  month = "Feb",
  year = "2010",
}

@article{zaharia2016apache,
  title={{Apache Spark}: a unified engine for big data processing},
  author={Zaharia, Matei and Xin, Reynold S and Wendell, Patrick and Das, Tathagata and Armbrust, Michael and Dave, Ankur and Meng, Xiangrui and Rosen, Josh and Venkataraman, Shivaram and Franklin, Michael J and others},
  journal={Communications of the ACM},
  volume={59},
  number={11},
  pages={56--65},
  year={2016},
  publisher={ACM}
}

@article{Ballintijn:2006ni,
    author = "Ballintijn, M. and Biskup, M. and Brun, R. and Ganis, G. and Kickinger, G. and Peters, A. and Rademakers, F. and Canal, P. and Feichtinger, D.",
    editor = "Blumlein, J. and Friebel, W. and Naumann, T. and Riemann, T. and Wegner, P. and Perret-Gallix, D.",
    title = "{Parallel interactive data analysis with PROOF}",
    doi = "10.1016/j.nima.2005.11.100",
    journal = "Nucl. Instrum. Meth. A",
    volume = "559",
    pages = "13--16",
    year = "2006"
}

@misc{htcondor,
author={{HTCondor Team}},
title={{DAGMan (Directed Acyclic Graph Manager)}},
url = "https://research.cs.wisc.edu/htcondor/dagman/dagman.html",
}

@inproceedings{rocklin2015dask,
  title={Dask: Parallel computation with blocked algorithms and task scheduling},
  author={Rocklin, Matthew},
  booktitle={Proceedings of the 14th python in science conference},
  number={130-136},
  year={2015},
  organization={Citeseer}
}

@article{Adamec:2021vkl,
    author = "Adamec, Matous and Attebury, Garhan and Bloom, Kenneth and Bockelman, Brian and Lundstedt, Carl and Shadura, Oksana and Thiltges, John",
    title = "{Coffea-casa: an analysis facility prototype}",
    eprint = "2103.01871",
    archivePrefix = "arXiv",
    primaryClass = "cs.DC",
    doi = "10.1051/epjconf/202125102061",
    journal = "EPJ Web Conf.",
    volume = "251",
    pages = "02061",
    year = "2021"
}

@article{Galewsky:2020xig,
    author = "Galewsky, B. and Gardner, R. and Gray, L. and Neubauer, M. and Pivarski, J. and Proffitt, M. and Vukotic, I. and Watts, G. and Weinberg, M.",
    editor = "Doglioni, C. and Kim, D. and Stewart, G. A. and Silvestris, L. and Jackson, P. and Kamleh, W.",
    title = "{ServiceX A Distributed, Caching, Columnar Data Delivery Service}",
    doi = "10.1051/epjconf/202024504043",
    journal = "EPJ Web Conf.",
    volume = "245",
    pages = "04043",
    year = "2020"
}

@article{lefevre:login20,
 author = {Jeff LeFevre and Carlos Maltzahn},
 journal = {USENIX ;login:},
 number = {2},
 title = {{SkyhookDM:} Data Processing in Ceph with Programmable Storage},
 volume = {45},
 year = {2020}
}

@inproceedings{lam2015numba,
  title={{Numba: A LLVM-based Python JIT compiler}},
  author={Lam, Siu Kwan and Pitrou, Antoine and Seibert, Stanley},
  booktitle={Proceedings of the Second Workshop on the LLVM Compiler Infrastructure in HPC},
  pages={1--6},
  year={2015}
}

@InProceedings{ henry_schreiner-proc-scipy-2020,
  author    = { {H}enry {S}chreiner and {H}ans {D}embinski and {S}huo {L}iu and {J}im {P}ivarski },
  title     = { {B}oost-histogram: {H}igh-{P}erformance {H}istograms as {O}bjects },
  booktitle = { {P}roceedings of the 19th {P}ython in {S}cience {C}onference },
  pages     = { 63 - 69 },
  year      = { 2020 },
  editor    = { {M}eghann {A}garwal and {C}hris {C}alloway and {D}illon {N}iederhut and {D}avid {S}hupe },
  doi       = { 10.25080/Majora-342d178e-009 }
}

@article{Dembinski:2020dic,
    author = "Dembinski, Hans Peter and Pivarski, Jim and Schreiner, Henry",
    editor = "Doglioni, C. and Kim, D. and Stewart, G. A. and Silvestris, L. and Jackson, P. and Kamleh, W.",
    title = "{Recent developments in histogram libraries}",
    doi = "10.1051/epjconf/202024505014",
    journal = "EPJ Web Conf.",
    volume = "245",
    pages = "05014",
    year = "2020"
}

@inproceedings{citeulike:363715,
  author = {Brun, Rene and Rademakers, Fons},
  booktitle = {AIHENP'96 Workshop, Lausane},
  howpublished = {http://root.cern.ch/},
  journal = {Nucl. Inst. \& Meth. in Phys. Res. A},
  pages = {81--86},
  title = {{ROOT - An Object Oriented Data Analysis Framework}},
  volume = 389,
  year = 1996
}

@article{Vasilev_2012,
	doi = {10.1088/1742-6596/396/5/052071},
	url = {https://doi.org/10.1088/1742-6596/396/5/052071},
	year = 2012,
	month = {dec},
	publisher = {{IOP} Publishing},
	volume = {396},
	number = {5},
	pages = {052071},
	author = {V Vasilev and Ph Canal and A Naumann and P Russo},
	title = {Cling {\textendash} The New Interactive Interpreter for {ROOT} 6},
	journal = {Journal of Physics: Conference Series}
}

@article{Lavrijsen:2015khc,
    author = "Lavrijsen, W.",
    title = "{Python in the Cling World}",
    doi = "10.1088/1742-6596/664/6/062029",
    journal = "J. Phys. Conf. Ser.",
    volume = "664",
    number = "6",
    pages = "062029",
    year = "2015"
}

@misc{compiler_research,
title={{Princeton/CERN Compiler Research Group}},
url = "https://compiler-research.org/",
}

@article{bezanson2017julia,
  title={Julia: A fresh approach to numerical computing},
  author={Bezanson, Jeff and Edelman, Alan and Karpinski, Stefan and Shah, Viral B},
  journal={SIAM review},
  volume={59},
  number={1},
  pages={65--98},
  year={2017},
  publisher={SIAM},
  url={https://doi.org/10.1137/141000671}
}

@article{Besjes:2015vns,
    author = "Besjes, G. J. and Baak, M. and C\^ot\'e, D. and Koutsman, A. and Lorenz, J. M. and Short, D.",
    title = "{HistFitter: a flexible framework for statistical data analysis}",
    doi = "10.1088/1742-6596/664/7/072004",
    journal = "J. Phys. Conf. Ser.",
    volume = "664",
    number = "7",
    pages = "072004",
    year = "2015"
}

@article{pyhf_joss,
  doi = {10.21105/joss.02823},
  url = {https://doi.org/10.21105/joss.02823},
  year = {2021},
  publisher = {The Open Journal},
  volume = {6},
  number = {58},
  pages = {2823},
  author = {Lukas Heinrich and Matthew Feickert and Giordon Stark and Kyle Cranmer},
  title = {pyhf: pure-Python implementation of HistFactory statistical models},
  journal = {Journal of Open Source Software}
}

@article{Cranmer:2012sba,
    author = "Cranmer, Kyle and Lewis, George and Moneta, Lorenzo and Shibata, Akira and Verkerke, Wouter",
    collaboration = "ROOT",
    title = "{HistFactory: A tool for creating statistical models for use with RooFit and RooStats}",
    reportNumber = "CERN-OPEN-2012-016",
    month = "6",
    year = "2012"
}

@techreport{CMS-NOTE-2011-005,
      title         = "{Procedure for the LHC Higgs boson search combination in
                       Summer 2011}",
      institution   = "CERN",
      collaboration = "The ATLAS Collaboration, The CMS Collaboration, The LHC
                       Higgs Combination Group",
      address       = "Geneva",
      reportNumber  = "CMS-NOTE-2011-005, ATL-PHYS-PUB-2011-11",
      month         = "Aug",
      year          = "2011",
      url           = "https://cds.cern.ch/record/1379837",
}

@proceedings{cranmer_kyle_2021_4627038,
  title        = {Building and steering template fits with cabinetry},
  year         = 2021,
  publisher    = {Zenodo},
  month        = mar,
  note         = {{This work was supported by the U.S. National 
                   Science Foundation (NSF) Cooperative Agreement
                   OAC-1836650 (IRIS-HEP).}},
  doi          = {10.5281/zenodo.4627038},
  url          = {https://doi.org/10.5281/zenodo.4627038}
}

@article{iminuit,
  author={Hans Dembinski and Piti Ongmongkolkul et al.},
  title={scikit-hep/iminuit},
  DOI={10.5281/zenodo.3949207},
  publisher={Zenodo},
  year={2020},
  month={Dec},
  url={https://doi.org/10.5281/zenodo.3949207}
}

@article{James:1975dr,
    author = "James, F. and Roos, M.",
    title = "{Minuit: A System for Function Minimization and Analysis of the Parameter Errors and Correlations}",
    reportNumber = "CERN-DD-75-20",
    doi = "10.1016/0010-4655(75)90039-9",
    journal = "Comput. Phys. Commun.",
    volume = "10",
    pages = "343--367",
    year = "1975"
}

@article{Eschle:2020ghu,
    author = "Eschle, Jonas and Puig, Albert Navarro and Silva Coutinho, Rafael and Serra, Nicola",
    editor = "Doglioni, C. and Kim, D. and Stewart, G. A. and Silvestris, L. and Jackson, P. and Kamleh, W.",
    title = "{{Zfit: scalable Pythonic fitting}}",
    doi = "10.1051/epjconf/202024506025",
    journal = "EPJ Web Conf.",
    volume = "245",
    pages = "06025",
    year = "2020"
}

@techreport{ATL-PHYS-PUB-2019-029,
      title         = "{Reproducing searches for new physics with the ATLAS
                       experiment through publication of full statistical
                       likelihoods}",
      institution   = "CERN",
      collaboration = "ATLAS Collaboration",
      address       = "Geneva",
      reportNumber  = "ATL-PHYS-PUB-2019-029",
      month         = "Aug",
      year          = "2019",
      url           = "https://cds.cern.ch/record/2684863",
}

@article{Rodrigues:2020syo,
    author = "Rodrigues, Eduardo and others",
    editor = "Doglioni, C. and Kim, D. and Stewart, G. A. and Silvestris, L. and Jackson, P. and Kamleh, W.",
    title = "{The Scikit HEP Project -- overview and prospects}",
    eprint = "2007.03577",
    archivePrefix = "arXiv",
    primaryClass = "physics.comp-ph",
    doi = "10.1051/epjconf/202024506028",
    journal = "EPJ Web Conf.",
    volume = "245",
    pages = "06028",
    year = "2020"
}

@misc{grand_challenge,
author = {{IRIS-HEP}},
title = {{Grand Challenges}},
url = "https://iris-hep.org/grand-challenges.html",
}

\end{document}